\documentclass[12pt]{iopart}
\usepackage{iopams} 
\usepackage{epsf}
\usepackage{graphicx}
\usepackage{subfigure}

\newcommand{\bogo}{Bogoyavlenskiy}

\begin{document}

\title{Is it really possible to grow isotropic on-lattice diffusion-limited aggregates?}

\author{S. G. Alves$\dagger$ and S. C. Ferreira Jr.$\ddagger$\footnote[3]{To whom correspondence should be addressed (silviojr@ufv.br).}}

\address{\small $\dagger$Departamento de F\'{\i}sica, Universidade Federal de Minas Gerais, CP 702, 30161-970, Belo Horizonte, MG, Brazil\\$\ddagger$Departamento de F\'{\i}sica, Universidade Federal Vi\c{c}osa, 36571-000, Vi\c{c}osa, MG, Brazil}

\begin{abstract}
In a recent paper (\bogo~V A 2002 \JPA \textbf{35} 2533), an algorithm aiming to generate isotropic clusters of the on-lattice diffusion-limited aggregation (DLA) model was proposed. The procedure consists of aggregation probabilities proportional to the squared number of occupied sites ($k^2$). In the present work, we analyzed this algorithm using the noise reduced version of the DLA model and large scale simulations. In the noiseless limit, instead of isotropic patterns, a $45^\circ$ ($30^\circ$) rotation in the anisotropy directions of the clusters grown on square (triangular) lattices was observed. A generalized algorithm, in which the aggregation probability is proportional to $k^\nu$, was proposed. The exponent $\nu$ has a nonuniversal critical value $\nu_c$, for which the patterns generated in the noiseless limit exhibit the original (axial) anisotropy for $\nu<\nu_c$ and the rotated one (diagonal) for $\nu>\nu_c$. The values $\nu_c = 1.395\pm0.005$ and $\nu_c = 0.82\pm 0.01$ were found for square and triangular lattices, respectively. Moreover, large scale simulations show that there are a nontrivial relation between noise reduction and anisotropy direction. The case $\nu=2$ (\bogo's rule) is an example where the patterns exhibit the axial anisotropy for small and the diagonal one for large noise reduction.
\end{abstract}




\section{Introduction}

Although proposed more than two decades ago, the diffusion-limited aggregation (DLA) model \cite{Witten} constitutes an actual and challenging theoretical problem. This simple nonequilibrium growth process generates fractal structures with nontrivial scaling, which was subject of recent studies \cite{MeakinBook,Somfai,Mandelbrot,FerreiraPRE}. In spite of its simplicity, the DLA model was related to several physical and biological applications, such as electrodeposition \cite{Matsushita}, viscous fingering \cite{Maloy}, bacterial colonies \cite{Matsushita2}, and neurite formation \cite{Caserta}. In the DLA model, particles released at points distant from the cluster execute random walks of unitary length. If a walker finds a site neighboring the cluster, it irreversibly sticks to this site. Both, on- and off-lattice versions of the model were intensively investigated \cite{Ball,Tolman,Kaufman}.

\begin{figure}[hbt]
\begin{center}
\epsfxsize=11.3cm
\epsfbox{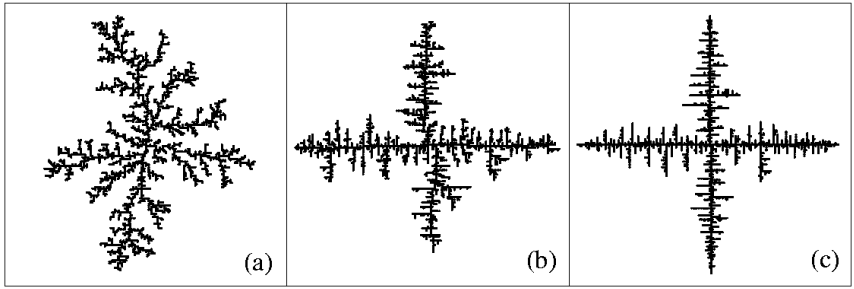}
\end{center}
\caption{DLA patterns grown on the square lattice with distinct noise reduction parameters: (a) $M=2^0$ (without noise reduction), (b) $M=2^2$, (c) $M=2^4$. The regions delimited by the boxes correspond to $200\times200$ lattice unities. The number of particles in the clusters varies from $1900$ ($M=2^4$) to $2800$ ($M=2^0$).}
\label{fig:dlawitten}
\end{figure}

The clusters generated by the on-lattice DLA model are unavoidable very sensitive to the lattice anisotropy \cite{Ball,Meakin1986}. The anisotropy is also a determinant factor present in other on-lattice growth processes as, for example, the Eden model \cite{Zabolitzky,Kertez,Batchelor}. A procedure widely used to expose cluster anisotropy is the noise reduction, which consists of a set of counters at each growth site that are increased by a unit every time these sites are visited. An empty site is occupied only after its selection for $M$ times. In figure \ref{fig:dlawitten}, clusters generated by  the original DLA model on the square lattice with distinct $M$ values are shown. One can see that the effects of the axial anisotropy are enhanced as the noise reduction parameter is increased. 

Recently Bogoyavlenskiy proposed an aggregation rule for the DLA model aiming to remove the anisotropy of clusters grown on square lattices \cite{BogoJPA}. We refer to the \bogo's algorithm for the DLA model as BDLA along the text. The rule is the following. When a walker reaches an empty site neighboring the cluster, it will stick to this site with a probability $P_k$ given by
\begin{equation}
P_1:P_2:P_3:\cdots:P_k=1^2:2^2:3^2\cdots k^2,
\end{equation}
where $k$ is the number of occupied neighbors of the walker. When the aggregation does not occur, the particle is excluded and a new walk begins. This idea was based on a previous work, in which was verified that an aggregation probability proportional to the squared mean density is a special rule which generates isotropic patterns in a quasicontinuum extension of the DLA model \cite{BogoPRE}. Deviations from this rule ($P\propto \rho^\alpha$ with $\alpha\ne 2$) result in anisotropic patterns with preferential growth along the axial or diagonal directions if $\alpha<2$ or $\alpha>2$, respectively. Three procedures were used to test the algorithm validity: (i) noise reduction algorithms with $M\le 2^4$; (ii) evaluation of the angular distributions of the particle density; and (iii) large scale simulations ($10^5$ particles). 

In the present work, we analyze BDLA clusters grown in square and triangular lattices using large noise reduction and large scale simulations. The computational procedures and the noiseless limit results are discussed in section \ref{sec2}. A generalized version of this algorithm is presented and discussed in section \ref{sec3}, and some conclusions are drawn in section \ref{sec4}.

\section{Noise reduction in the Bogoyavlenskiy's algorithm}
\label{sec2}

\begin{figure}[hbt]
\begin{center}
\epsfxsize=15cm
\epsfbox{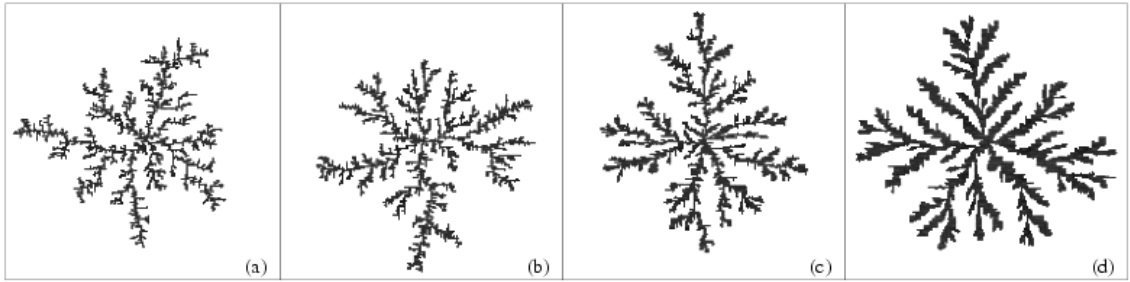}
\end{center}
\caption{BDLA clusters grown on square lattices using distinct noise reduction parameters: (a) $M=2^0$ (without noise reduction), (b) $M=2^1$, (c) $M=2^2$, and (d) $M=2^4$. The regions delimited by the boxes correspond to $200$ lattice unities and the number of particles varies from 3000 ($M=2^0$) to 7000 ($M=2^4$). These noise reduction parameters are the same used in reference \cite{BogoJPA}.}
\label{fig:dlabogo}
\end{figure}

In the original model proposed by Witten and Sander \cite{Witten}, the particles are released, one at a time, at points distant from the cluster and execute random walks on a square lattice. If a particle reaches an empty site neighboring the aggregate\footnote[1]{At the beginning of the simulation, a single particle at the center of the lattice constitutes the cluster.}, it irreversibly sticks to this site. The distance between the center of the lattice and the hypersphere in which the walkers are released can be chosen as few lattices units larger than  $R_0$, the maximum distance from the center of the lattice of a particle lying on the aggregate. Moreover, if the particle is too far way from the cluster, it is excluded and a new one released at the launching hypersphere. We used the launching and killing radius equal to $R_l=R_0+5$ and $R_k=100R_0$, respectively. Since this algorithm is very inefficient, the analysis with large noise reduction becomes prohibitive even for small systems. Therefore, we used a standard method in which the particles are allowed to execute long steps if they are far way from the clusters \cite{Ball,Tolman,MeakinJPA,Ferreira}. 

\begin{figure}[hbt]
\begin{center}
\epsfxsize=11.3cm
\epsfbox{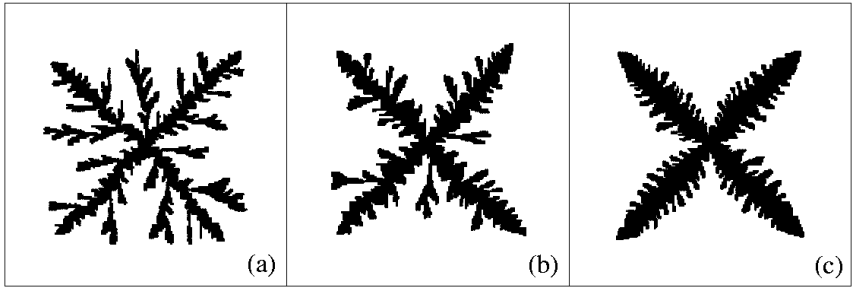}
\end{center}
\caption{BDLA clusters grown on square lattices using large noise reduction. The noise reduction parameters are (a) $M=2^5$, (b) $M=2^6$, and (c) $M=2^8$ and the regions delimited by the boxes correspond to $200\times200$ lattice unities.}
\label{fig:dlabogo2}
\end{figure}

For BDLA clusters grown on the square lattice, the aggregation probabilities are given by  $P_k=k^2/3^2$, where $k=1,2, \mbox{~or~}3$ is the number of occupied neighbors. Patterns corresponding to distinct noise reduction parameters $M$ are shown in figure \ref{fig:dlabogo}. These patterns suggest the absence of anisotropy, since the cluster with the largest noise reduction ($M=2^4$) does not exhibit preferential growth directions. Indeed, this can be reinforced comparing figure \ref{fig:dlabogo}(d) with the corresponding one for the original DLA model shown in figure \ref{fig:dlawitten}(c). Only the capillary width of these patterns, which enlarges with $M$, was modified. However, the noise reduction parameters used to generate these patterns are relatively small. In figure \ref{fig:dlabogo2}, simulations with larger $M$ values are shown. For $M=2^5$, one can clearly observe that the diagonal directions have preferential growth. This becomes more evident as larger $M$ values are used. In the noiseless limit, a regular fourfold structure with a $45^\circ$ rotation in relation to the axial directions was found. Therefore, grounded in the later results one can conclude that the \bogo's~algorithm does not remove the lattice anisotropy. Instead, it only causes a rotation of the anisotropy direction. 

The physical origin of this rotation can be understood through the following argument. In square lattices, the net effect of the anisotropy is to increase the growth probabilities in the axial directions. Using an aggregation probability increasing with the number of occupied neighbor sites, the growth of the sites with two or three occupied neighbors is favored. Due to the ramified morphology of DLA clusters, empty sites with three occupied neighbors are much more rare than those with one or two. Thus, the effective result of this rule is to increase the growth probability of sites with two occupied neighbors and, consequently, to increase the growth probabilities in the diagonal directions. If the growth probabilities of empty sites with two occupied neighbors become larger than those of the sites with one occupied neighbor, the rotation of the anisotropy direction emerges. 

\begin{figure}[hbt]
\begin{center}
\epsfxsize=15cm
\epsfbox{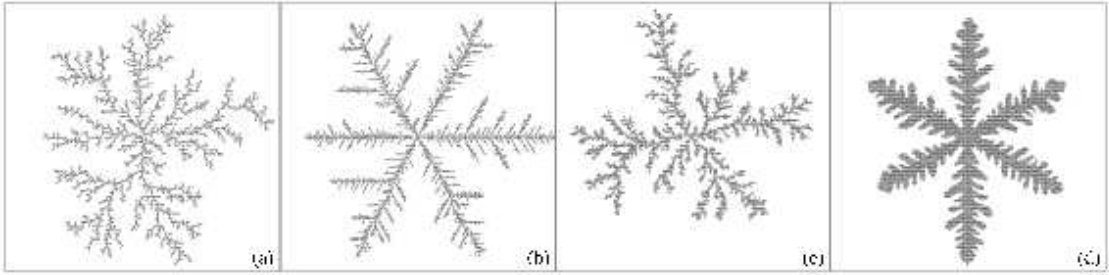}
\end{center}
\caption{DLA((a) and (b)) and BDLA ((c) and (d)) clusters grown on triangular lattices. The noise reduction parameter $M=2$ was used in (a) and (c), while $M=2^8$ was used in (b) and (d). The boxes correspond to $200\times200$ lattice unities. }
\label{fig:hex}
\end{figure}

Another important feature of the BDLA model is the dependence on the lattice structure of the anisotropy rotations, as one can observe in figure \ref{fig:hex}, where clusters grown on triangular lattices are shown. The aggregation probabilities for BDLA are given by
\begin{equation}
P_k=k^2/5^2\mbox{~where~} k=1,\cdots,5.
\end{equation}
In figures \ref{fig:hex}(a) and (b) we show patterns generated by the original DLA model for $M=2$ and $M=2^8$. In turn, figures \ref{fig:hex}(c) and (d) illustrate the corresponding patterns of the BDLA model. Similarly to the results of the square lattice, one can see that the effects of the algorithm are the enlargement of the capillary width and a $30^\circ$ rotation in the anisotropy directions. In the noiseless limit, the patterns are characterized by a regular sixfold structure in analogy with the square lattice case.

\begin{figure}[hbt]
\begin{center}
\epsfxsize=15cm
\epsfbox{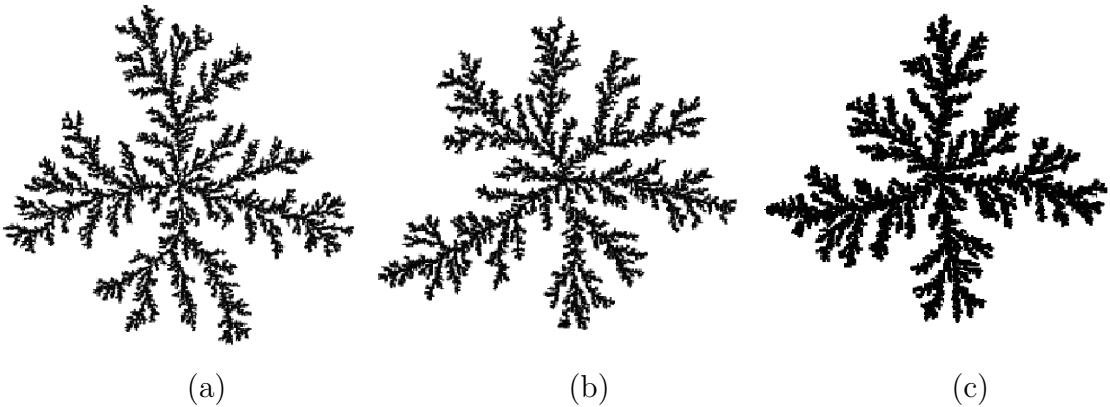}
(a)~~~~~~~~~~~~~~~~~~~~~~~~~~~~~~~~~(b)~~~~~~~~~~~~~~~~~~~~~~~~~~~~~~~~~(c)
\end{center}
\caption{(a) Clusters with $10^5$ particles generated by the DLA model and clusters generate by the BDLA model with (b) $10^5$ and (c) $10^6$ particles. These clusters were grown in lattices $2000\times2000$ in (a) and (b), and $7000\times7000$ in (c).}
\label{fig:large}
\end{figure}

In order to confirm the results predicted by the large noise reduction analysis, we extend the large scale simulation performed by \bogo~\cite{BogoJPA}. In figure \ref{fig:large}, large aggregates generated using the DLA and BDLA models are shown. In figures \ref{fig:large}(a) and (b), we reproduced the simulations shown in reference \cite{BogoJPA}, in which clusters containing $10^5$ particles were generated for both rules. In figure \ref{fig:large}(a), where the original model was used, the effects of the anisotropy are neat but, they do not appear in figure \ref{fig:large}(b). Thus, these figures suggest the anisotropy removal. However, if we continue the simulation shown in figure \ref{fig:large}(b) up to the cluster reaches $10^6$ particles, the axial anisotropy emerges, as one can see in figure \ref{fig:large}(c).

\section{The generalized algorithm}
\label{sec3}

The distinct geometries observed for the noiseless DLA clusters grown using or not the \bogo's~rule suggest that this procedure can be generalized. Therefore, we modified the algorithm by defining an aggregation probability proportional to a power of the number of occupied neighbors, i.e.,
\begin{equation}
P_k=\left(\frac{k}{n}\right)^\nu\mbox{~~where~}k=1,\cdots,n.
\end{equation}
Here, $n$ is the lattice coordination number and $\nu$ a generalization parameter. For $\nu=0$ we recover the original DLA model and $\nu=2$ the BDLA. Therefore, one expects a critical value $\nu_c\in(0,2)$ delimiting these regimes. This aggregation rule is similar to that used in a modified DLA model investigated by Batchelor and Henry \cite{BatchelorH}. In this model, which was studied in square lattices, the aggregation probabilities are given by $P_k=r^{3-k}$, where $r\in(0,1]$ is a parameter. The noiseless limit of this model reveals a rich variety of morphologies that obey a nontrivial relation with the $r$ parameter. Notice that in this aggregation rule the number of occupied neighbors appears in the power, contrasting the present generalization, in which this number appears in the basis.

\begin{figure}[h]
\begin{center}
\epsfxsize=11.3cm
\epsfbox{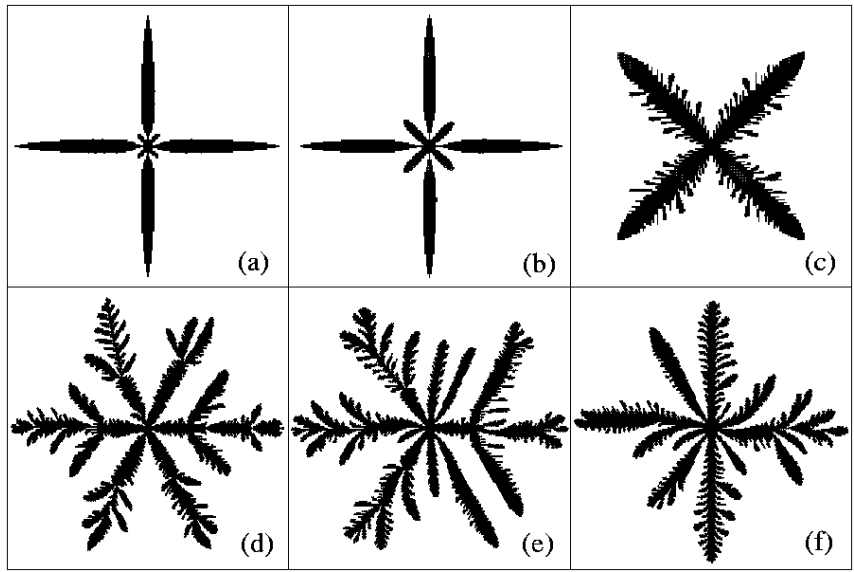}
\end{center}
\caption{DLA clusters grown using the generalized algorithm. The noise reduction parameters are equal to $M=2^{13}$ and $M=2^{17}$ for square ((a)-(c)) and triangular ((d)-(f)) lattices, respectively. The generalization parameters are (a) $\nu=1.10$, (b) $\nu=1.36$, (c) $\nu=1.40$, (d) $\nu=0.77$, (e) $\nu=0.80$, and (f) $\nu=0.83$. The boxes correspond to $200\times200$ lattice unities.}
\label{fig:nus}
\end{figure}

Figures \ref{fig:nus}(a)-(c) show clusters generated in square lattices for three values of the parameter $\nu$. Only the limit of large noise reduction was considered in this figure. For small $\nu$ values, the morphologies are ruled by the usual square lattice anisotropy. When the $\nu$ values are increased, the clusters exhibit small diagonal modes coexisting with the fourfold patterns (e.g., figure 1(a)). As the $\nu$ exponent is increased, the diagonal modes become more neat (e.g., figure 1(b)), up to a critical value $\nu_c$, when the diagonal anisotropy suddenly becomes the rule (figure 1(c)). Simulations with  $M=2^{15}$ (not shown in figure \ref{fig:nus}) show that the critical value of $\nu$ is in the interval $1.39<\nu_c<1.40$ and, therefore, the best estimate which can be made about the exponent value is $\nu_c=1.395\pm0.005$. The influence of the $\nu$ parameter in the morphology of clusters grown on triangular lattices is illustrated in figures \ref{fig:nus}(d)-(f). Even using a noise reduction 16 times larger than that used for figures \ref{fig:nus}(a)-(c) ($M=2^{17}$), these patterns do not display the closely regular structures observed for the square lattices. These simulations provided a critical $\nu$ parameter in the range $0.81<\nu_c<0.83$ implying that the best estimate of the critical parameter for triangular lattices is $\nu_c=0.82 \pm 0.01$.

\begin{figure}[hbt]
\begin{minipage}{1.8cm}
~\\
\bigskip\bigskipamount1.7cm
~\\
\bigskip\bigskipamount1.7cm
$\nu=4.0$\\
\bigskip\bigskipamount1.7cm
$\nu=3.0$\\
\bigskip\bigskipamount1.7cm
$\nu=2.5$\\
\bigskip\bigskipamount1.7cm
$\nu=2.0$\\
\bigskip\bigskipamount1.7cm
$\nu=1.8$\\
\bigskip\bigskipamount1.7cm
$\nu=1.0$
\end{minipage}
\begin{minipage}{14.2cm}
\includegraphics[clip=true,width=14.2cm]{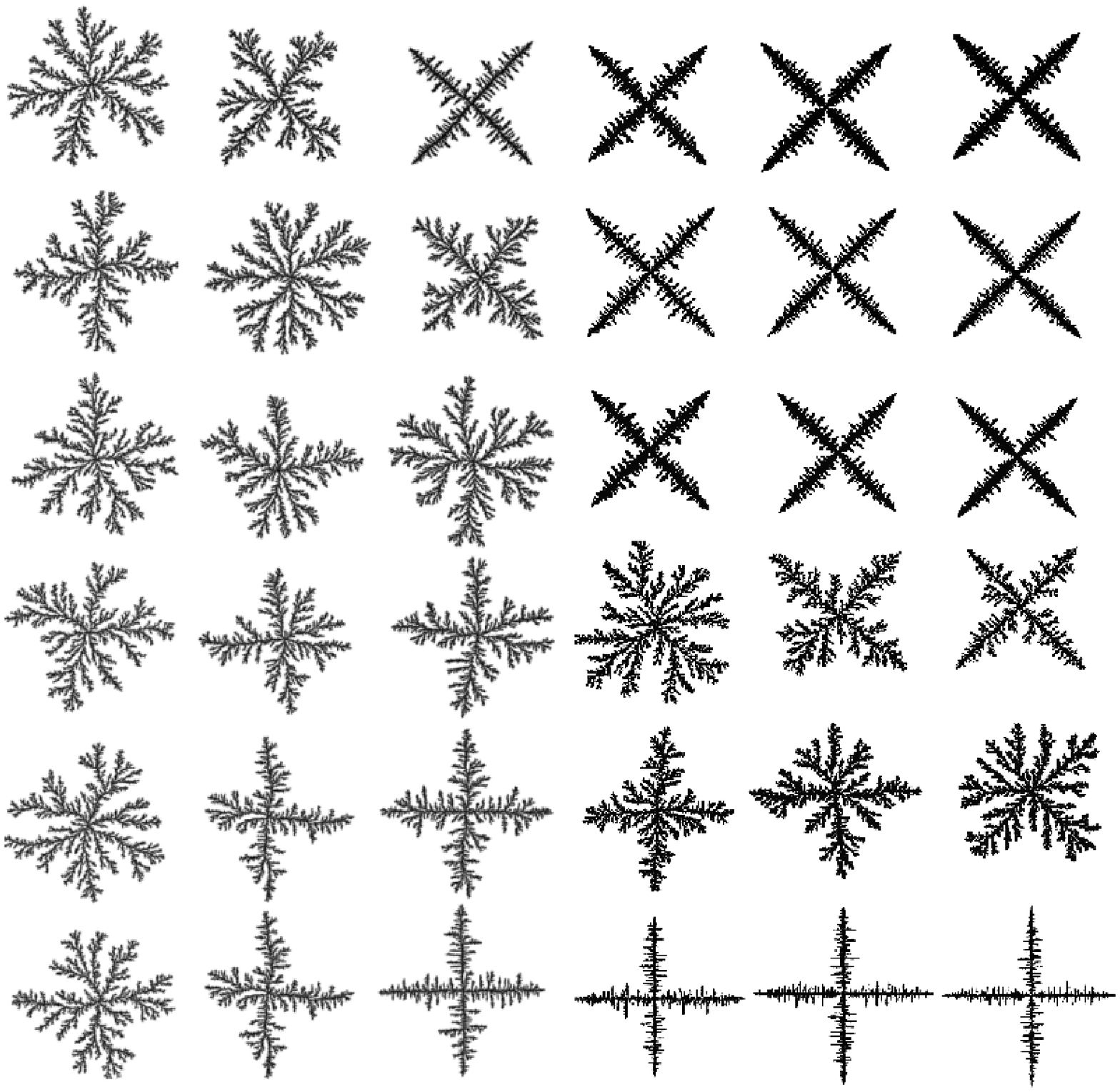}\\
\hspace*{0.55cm}$M=2^0\;\;\;\;\;\;\;\;M=2^1\;\;\;\;\;\;\;\;M=2^2\;\;\;\;\;\;\;\;M=2^4\;\;\;\;\;\;\;\;M=2^5\;\;\;\;\;\;\;\;M=2^6$
\end{minipage}
\caption{Morphological diagram for the generalized BDLA model using square lattices with $2000\times2000$ sites. The corresponding $\nu$ and $M$ values are shown at the left of the rows and at the bottom of the columns, respectively. The numbers of particles for these aggregates are of the order of $10^5$. }
\label{fig:diagram}
\end{figure}

The relation between the cluster anisotropy and the parameters $\nu$ and $M$ is nontrivial, as indicated in the morphological diagram shown in figure \ref{fig:diagram}. These clusters were generated with lattices of size $2000\times2000$ and several values of $\nu$ and $M$. It is clear from this diagram that for small $\nu$ values the axial anisotropy is dominant while the diagonal one rules the patterns for large $\nu$. However, the dependence on the noise reduction is more complex. For $\nu<\nu_c=1.395\pm 0.05$, the noise reduction acts like in the DLA model, i.e., the larger the noise reduction more evident the axial anisotropy. Similarly, for $\nu$ above an upper threshold $\nu^{*}\approx 3.4\pm 0.2$, the noise reduction always enhances the diagonal anisotropy. But, for intermediate $\nu$ values, the noise reduction enhances the axial anisotropy for small $M$ while the diagonal one becomes the rule for large $M$. This is the case for $\nu=2$ which corresponds to the BDLA model.  In particular, the largest noise reduction used in the reference \cite{BogoJPA}, $M=2^4$, corresponds to a point near to the region of transition. In this region, both axial and diagonal anisotropies have approximately the same weight, and the patterns are apparently not sensitive to the lattice anisotropy.
\begin{figure}[hbt]
\begin{center}
\subfigure[]{\epsfxsize=7.5cm\epsfbox{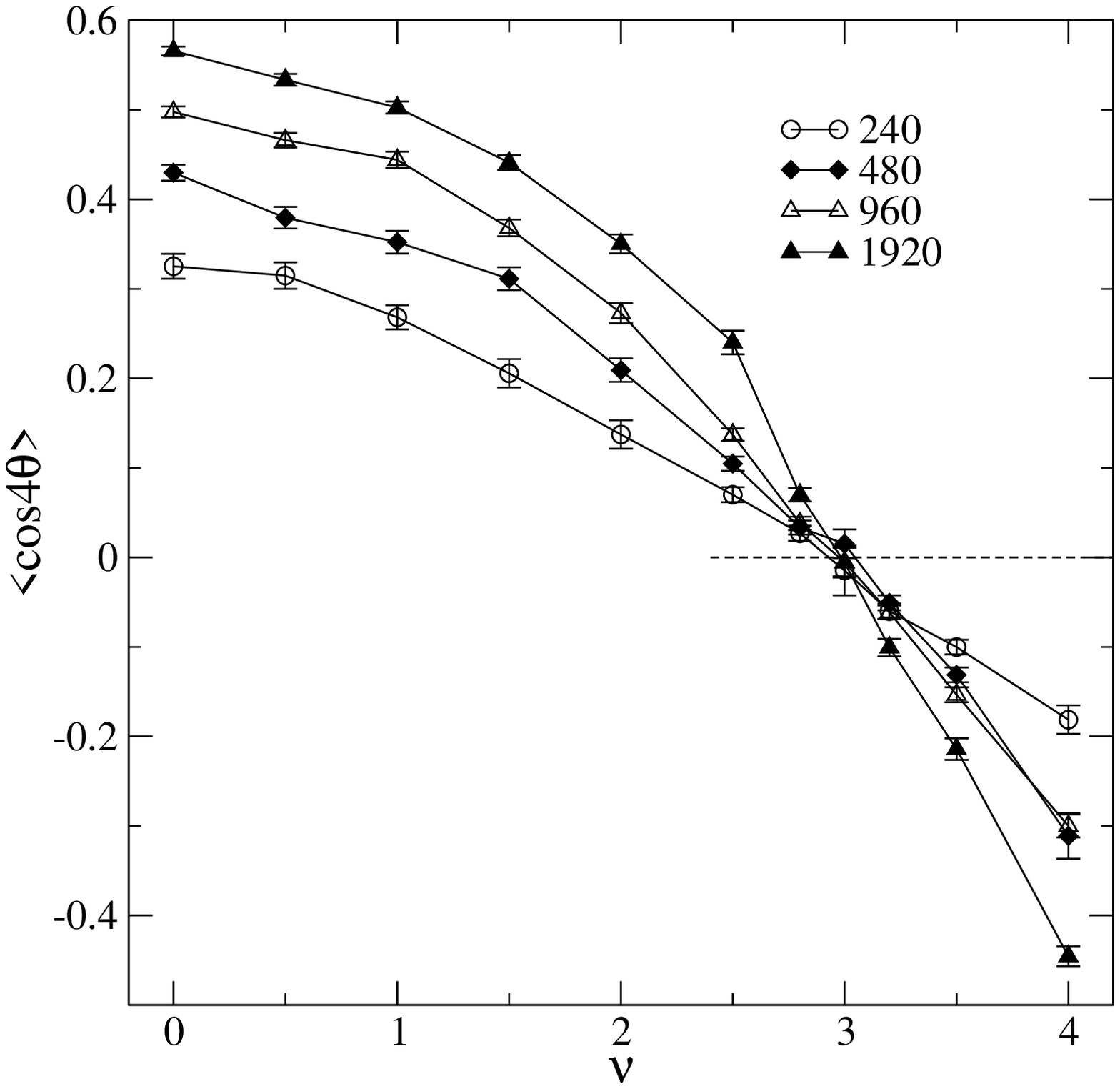}}
\subfigure[]{\epsfxsize=7.5cm\epsfbox{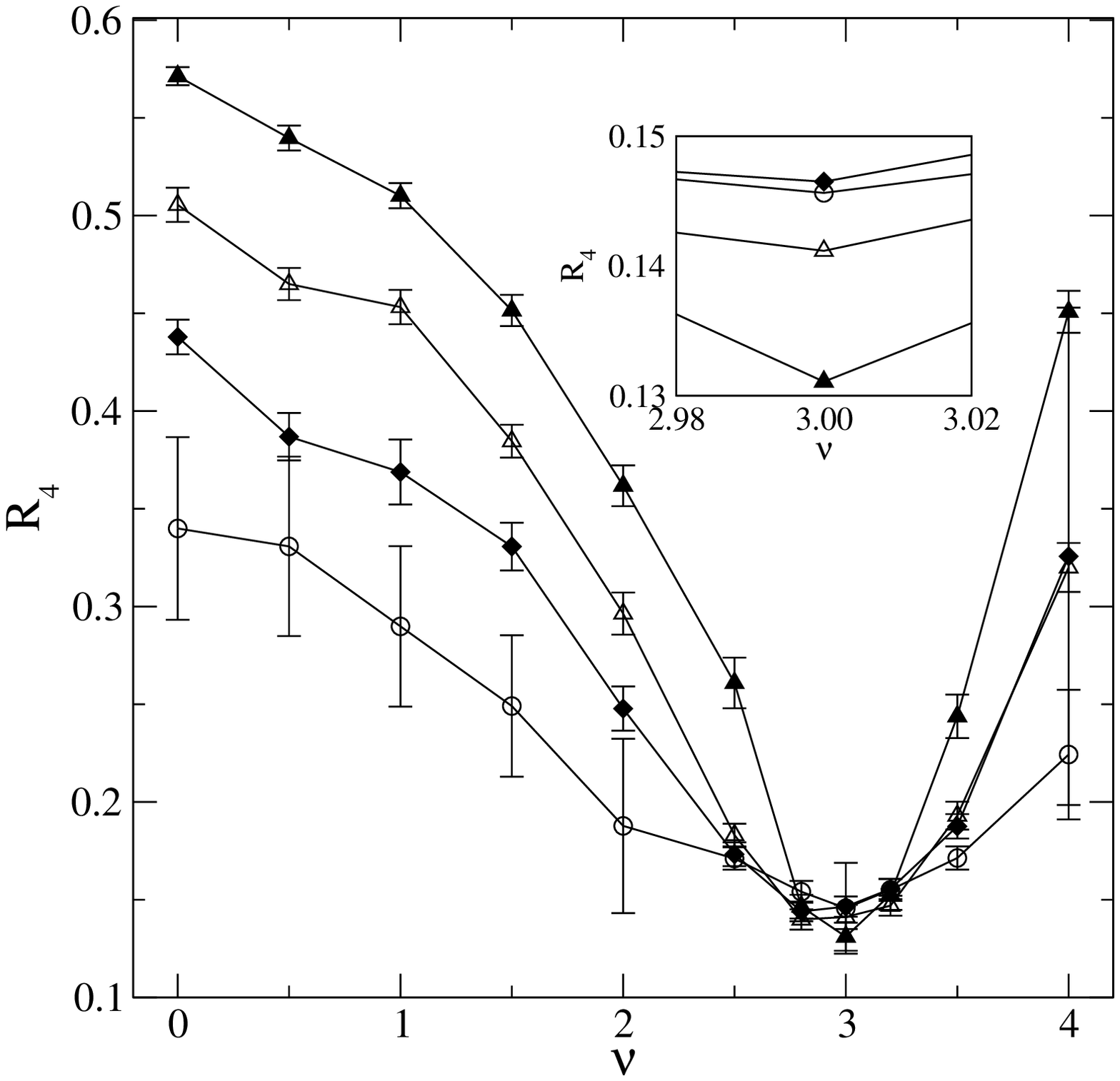}}
\subfigure[]{\epsfxsize=7.5cm\epsfbox{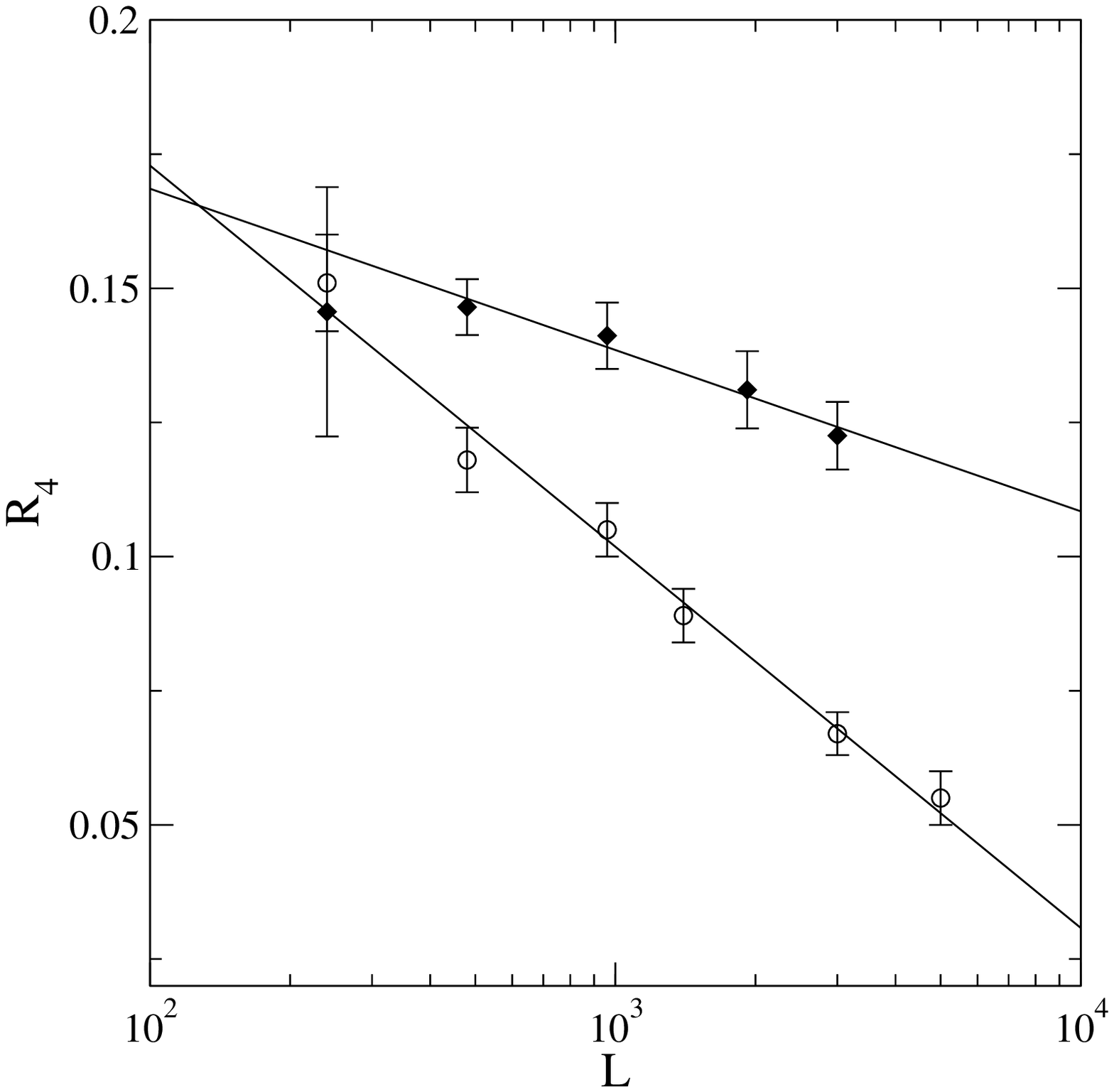}}
\end{center}
\caption{\label{fig:aniso} Measures of the anisotropies as functions of $\nu$ for distinct system sizes and noise reduction $M=2$. These quantities were defined in the text. In all simulations, the averages were done over 50 or 200 independent samples for $|\nu-\nu^\prime|> 0.5$ or  $|\nu-\nu^\prime|\le 0.5$, respectively. Figure \ref{fig:aniso}(c) shows $R_4$ as function of the system size for the off-lattice DLA model ($\circ$) and the generalized model at $\nu=\nu^\prime$ ($\blacklozenge$) with the respective logarithm regressions. The inset of (b) shows a zoom around the minima of the curves $R_4$ versus $\nu$.}
\end{figure}

The determination of the transition line requires a quantitative measure of the anisotropy. Due to the fourfold symmetry of the patterns grown on square lattices, the fourth circular harmonic \cite{Ball,Ferreira}
\begin{equation}
\langle \cos 4\theta \rangle=\frac{1}{N}\sum_{n=1}^N\cos 4\theta_n,
\end{equation}
where the sum is done over all particles of the cluster and $\theta_n$ is the angular position measured in relation to the origin, is an useful quantity since $\langle \cos 4\theta \rangle > 0$ for patterns with axial anisotropy, $\langle \cos 4\theta \rangle < 0$ for patterns with diagonal anisotropy,  and $\langle \cos 4\theta \rangle = 0$ for isotropic or eightfold\footnote{Axial and diagonal anisotropy are present with the same weight.} patterns. Also, we evaluated the Fourier transform of the mass distribution defined by \cite{Moukarzel}
\begin{equation}
C_k=\left|\sum_{n=1}^N\mathrm{e}^{ik\theta_n}\right|.
\end{equation}
Again due to the symmetries of the system, we are only interested in the ratio $R_4=C_4/C_0$. In figures \ref{fig:aniso}(a) and (b), we show $\langle\cos4\theta\rangle$ and $R_4$ as functions of the parameter $\nu$ for distinct system sizes and noise reduction $M=2$. As expected, $\langle\cos4\theta\rangle$ is positive for small $\nu$ values and negative for large values, vanishing for some value $\nu^\prime$. The accuracy of these data allows us to estimate $\nu^\prime=3.0\pm0.1$ which, in this accuracy limit, is independent of the system size. The curves $R_4$ against $\nu$ exhibit positive minima at $\nu=\nu^\prime$ that, within the error bars, are approximately independent of the system size. For $\nu$ distant from $\nu^\prime$ ($|\nu-\nu^\prime|\gnapprox 0$), $R_4$ and $|\langle \cos 4\theta \rangle|$ are increasing functions of the system size for a fixed  $\nu$, demonstrating the axial ($\nu<\nu^\prime$) or diagonal ($\nu>\nu^\prime$) anisotropy enhancement. However, in inset of figure \ref{fig:aniso}(b), the zoom around the minima shows that $R_4(\nu\approx \nu^\prime)$ slowly decreases with the system size. So, it is important to compare the minima of these curves with the $R_4$ values obtained for off-lattice DLA clusters. For this task, we simulated $100$ off-lattice DLA clusters of linear sizes $240,~480,~960,~1400,~3000$ and $5000$ and the results are shown in figure \ref{fig:aniso}(c). As one can see, $R_4$ decreases logarithmically slow even for the isotropic off-lattice DLA clusters. In turn, one expects that $R_4\approx 0 $ for homogeneous structures, but this is not the case for DLA clusters, justifying the very low convergence to the asymptotic value. Also, figure \ref{fig:aniso}(c) shows the minima of the curves in figure \ref{fig:aniso}(b) as a function of the system size. Again, the data suggest a logarithmic decay, but it is slower than that obtained for off-lattice DLA-model. This very slow decay indicates that the cluster anisotropy is minimal for $\nu=\nu^\prime$, probably reflecting their eightfold structures. Notice that we cannot claim that the patterns could be isotropic since $R_4$ is also null for exactly eightfold morphologies. Finally, we repeated the previous procedure for other noise reduction parameters and found $\nu^\prime=3.4\pm0.2$, $2.35\pm 0.05$, $1.92\pm0.02$, and $1.79\pm0.01$ for  $M=1,~4,~16$ and $32$, respectively. 

\begin{figure}[hbt]
\begin{center}
\epsfxsize=13cm
\epsfbox{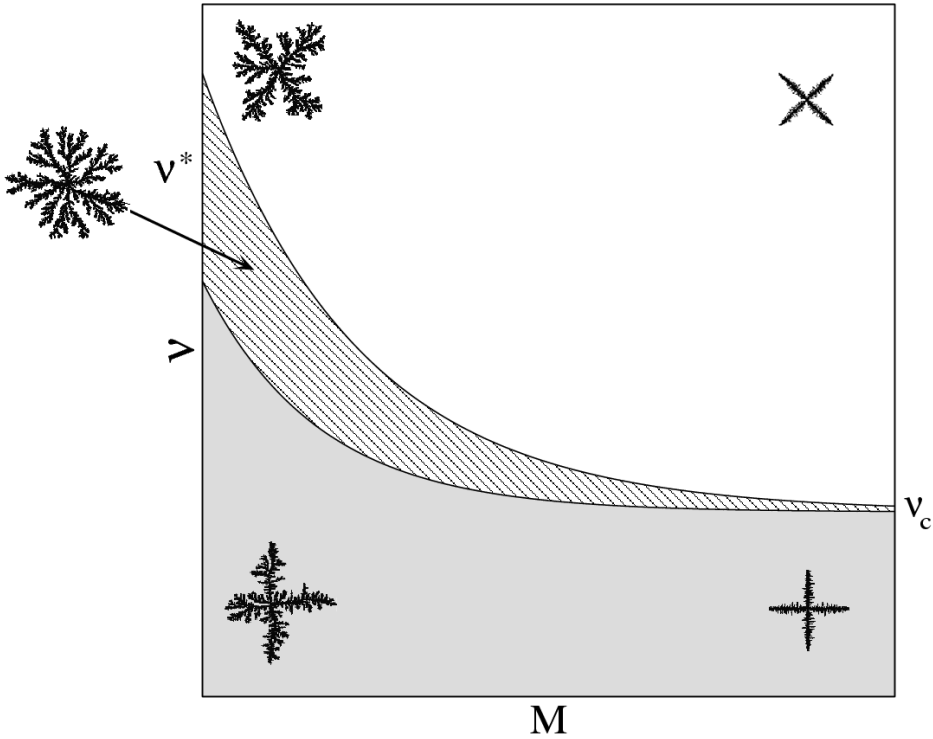}
\end{center}
\caption{Schematic representation of the morphological diagram. The region depicted in gray (white) is ruled by the axial (diagonal) anisotropy. The hachured region represents the uncertainty around the line delimiting the two regions of the morphological diagram. Typical morphologies corresponding to distinct regions of the morphological diagram are illustrated.}
\label{fig:schema}
\end{figure}

Baker and Ball \cite{Baker} studied the anisotropy imposed by the square lattice in a DLA model where, beside the sticking to the nearest neighbors of the aggregate, the particles  can also stick to the next nearest neighbors with a probability $p$. The parameter $p$ controls the direction of the anisotropy, i.e., for $p=0$ the patterns exhibit the axial anisotropy and for $p=1$ they exhibit the diagonal one. In this work, through real-space renormalization group, they found two stable and one unstable fixed points in the flux diagram $p-M$. The stable points $(p=0,M=\infty)$ and $(p=1,M=\infty)$ represent the axial and diagonal anisotropies, respectively, while the unstable one leads to eightfold structures. These results are in analogy with our generalization, which also seems to have two fixed stable points $(\nu=0,M=\infty)$ and $(\nu=\infty,M=\infty)$. Moreover, the qualitative morphological diagram and the quantitative characterization of the anisotropy suggest two distinct regions, in which only one of the anisotropy directions is present. The accuracy of the present simulations does not allow us to define the exact line delimiting these regions. Therefore, renormalization group analysis will be necessary to determine the exact flux diagram. In figure \ref{fig:schema}, we show a schematic representation of the morphological diagram, in which the increase of the $\nu^\prime$ error bars as $M\rightarrow 1$ is illustrated by the hachured region.

\section{Conclusions}
\label{sec4}

In this work we revisited a procedure used to simulate isotropic on-lattice DLA clusters. In this procedure, the aggregation probabilities are proportional to squared number of occupied neighbors, i.e., $P_k\propto k^2$. Contrasting the claims of this work, we demonstrate that this procedure implies in rotations of the anisotropy directions. More precisely, we found a $45^\circ$ and $30^\circ$ rotations for clusters grown on square and triangular lattices, respectively. Also, we studied a generalization of this method, in which $P_k\propto k^\nu$. The exponent $\nu$ has a nonuniversal critical value $\nu_c$ for which the patterns exhibit the usual (rotated) anisotropy directions for $\nu<\nu_c$ ($\nu>\nu_c$) in the noiseless limit. The values obtained for the square and triangular lattices were $\nu_c=1.395\pm 0.005$ and $\nu_c=0.82\pm0.01$, respectively.

The hypothesis of the anisotropy rotations was confirmed through large scale simulations of clusters grown using square lattices. In turn, we found a nontrivial relation between cluster anisotropy and the parameters related to the generalizations and to the noise reduction. To be more precise, for intermediate $\nu$ values, small noise reduction enhances the axial anisotropy while large noise reduction enhances the diagonal one. Also, using a quantitative analysis of the anisotropy, we determined the crossover value $\nu^\prime$ between axial and diagonal anisotropies for distinct noise reduction. In particular, for the case without noise reduction, we found $\nu^\prime=3.4\pm 0.2$. It is important to mention that all implementation details were varied as, for example, the random number generators and the use or not of the optimizations. In all cases, the same results were obtained. Thus, our final concluding remark is that the possibility of growing isotropic on-lattice DLA clusters has not been demonstrated yet. In the best scenario, one can mask the anisotropy effects including additional anisotropy growth directions. As a last comment, we would like to mention that, concerning the computer time, the simulations of the on-lattice generalized model at the point of minimal anisotropy ($\nu\approx\nu^\prime$) are slower than those for the off-lattice DLA model.

~\\~\\
\textbf{Acknowledgments}\\ We thank to M. L. Martins and J. G. Moreira by the critical reading of the manuscript. This work was supported by CNPq, CAPES, and FAPEMIG Brazilian agencies. 
~\\~\\

\noindent\textbf{References}\\

\end{document}